\documentclass[a4paper, 12pt]{article}
\usepackage{arxiv}
\pdfoutput = 1
\usepackage{tikz}
\usepackage{amsmath}
\usepackage{hyperref}       
\usepackage{url}
\hypersetup{
    colorlinks=true,
    linkcolor=blue,
    citecolor=red,
    filecolor=blue,      
    urlcolor=blue,}         
\usepackage{booktabs}       
\usepackage{amsfonts}       
\usepackage{nicefrac}       
\usepackage{lipsum}
\usepackage{graphics}
\usepackage{color,pxfonts,fix-cm}
\usepackage{latexsym}
\usepackage[mathletters]{ucs}
\usepackage[T1]{fontenc}
\usepackage[utf8x]{inputenc}
\usepackage{pict2e}
\usepackage{wasysym}
\usepackage{caption}
\usepackage[english]{babel}
\usepackage{ulem}
\usepackage{relsize}
\usepackage{cancel}
\everymath{\textstyle}
\usepackage{authblk}

\usepackage{mathtools}
\begin{document}
\renewcommand\Affilfont{\fontsize{11}{10}\itshape}

\title{\textbf{Study of the Curvature of Liquid Surface surrounding a Rotating Spherical Object in Gravity Free Space}}
\author{
  Rajdeep Tah$^1$\thanks{ rajdeep.tah@niser.ac.in},~ Sarbajit Mazumdar$^1$\thanks{ sarbajit.mazumdar@niser.ac.in} ~and Krishna Kant Parida $^1$\thanks{ krishna.parida@niser.ac.in}\\
 $^1$School of Physical Sciences\\
 $^1$ National Institute of Science Education and Research, Bhubaneswar, HBNI, P.O. Jatni, Khurda-752050, Odisha, India.\\
}
\maketitle
\begin{abstract}
Concept of curvature of liquid surrounding a spherical surface seems obvious in daily life, but based on earthly conditions everywhere. However, our understanding about the concept seems more transparent when we keep the system out of the usual earthly condition i.e. without gravity. Although existence of forces like adhesion and cohesion along liquid surface come to the fore even in the presence of other force like gravitational ones, but without gravity these forces are solely responsible for kind of observable phenomenon. Also, we introduced a form of force responsible for providing a form of potential dominating over the gravitational one. The discussion was provided an ingenious approach, by conserving surface energy, it still explains a lot about what can be done more to explore other properties of rotating liquids in free space. 
\end{abstract}
\section{Introduction}
Whenever we think of Curvature of a surface associated to a system, the first thing which comes to our mind is that it is related to the profile of the surface i.e. the physical state of the system's surface. In mathematical sense, the Curvature is defined as the amount by which a curve deviates from being a straight line or the amount by which a surface deviates from being a plane. The Curvature of curves drawn on a surface is the main tool for the defining and studying of the Curvature of the surface. Now, in our case we tend to study the Curvature of liquid surrounding a spherical surface which is in-turn rotating at an angular velocity $\omega$ but in the absence of a major factor i.e. Gravity. The absence of gravity in our case is a key factor which helps us to study our system in a more transparent and effective way i.e. it can be used to study the Curvature properties of fluids in gravity-free space. Although we are neglecting Gravity in our system, we will still consider the existence of other factors (like Cohesion forces; Adhesion forces; etc) which contribute for the change in the Curvature of the rotating system's surface. In the absence of gravity, this factors will be solely responsible for the change in the properties of the system. Further, we proceed to calculate the restriction on the angular velocity beyond which the centrifugal force will tend to spill the fluid out from the system and following it, we will try to study the various other properties of the system by conserving the Surface Energy through a much more ingenious approach.
\subsection*{\underline{Structure:}}
In Section \ref{descrip} we will describe our system by calculating the Equatorial Bulge ($n$) and also by deriving the Equation of the Spheroid formed by the fluid due to the rotation of the system. Then in Section \ref{ppcf} we will derive the expression for the Central Fluid Pressure ($P_{0}$) and following it we will derive the expression for the Containing Force ($F(x,y,z)$) of the system. After deriving both the previously mentioned expressions, we will derive the expression for the Angular Velocity of the system after finding out the Moment of Inertia ($I_{net}$) and Surface Area ($S^{'}$) of the system in Section \ref{av}. Then in Section \ref{lc} we will find the expressions for the Containing Force ($F(x)$) along $X$-axis, Limiting $\omega$ and the Containing Potential ($\Psi(x)$) along $X$-axis respectively which we are accounting under restricted conditions. After that in Section \ref{con} we will conclude our results which we inferred from the previous sections using various concepts followed by Acknowledgement in Section \ref{ack} and References respectively. 
\section{Description Of System}
\label{descrip}
We are considering a metal sphere is present in free space, with no net force on the system, and a liquid layer had completely engulfed the metal ball. The system is an isolated one, with no presence or effect of gravity and standard atmospheric pressure of $1$ $atm$. The physical conditions of the system can be compared to that present inside the International space station where habitable conditions are present with the exception of gravity.. Now the complete system is rotated with a constant angular velocity $\omega$ and was allowed to rotate with the same angular velocity.

We can say that many forces are acting within the system and primary one will be the gravitational force between the metal ball and the liquid layer. The system seems more or like a planet with a solid metallic core and a liquid atmosphere. But we have taken effect to to gravitational forces to be very negligible, since the mass of the metal ball as well as that of the liquid layer is very small. Also the Gravitational constant $G$ is of the value $6.674 \times 10^{-11}$ $m^3 kg^{-1} s^{-2}$, which is too small to make the gravitational force effective for calculations. Hence we have to take adhesion forces between the liquid particle and metal, and cohesion forces within the liquid particles to be main cause for keeping the complete system intact.

Let the radius of the metal ball be $R$ and the initial height of the liquid layer is taken to be $m$. Now the system is made up of material of different densities. Let the density of the liquid layer surrounding it be $\rho$ and that of the metal ball be $\rho_m$ respectively. Now from the initial state, since the complete system is being rotated, then the shape of the metal ball won't change, but the shape of the liquid layer will change due to non-rigid properties of it. Due to presence of centrifugal force, the liquid will try to flatten out in the space along equatorial plane, resulting in the formation of a spheroid shaped drop of liquid surrounding the spherical metal ball.
\subsection{Calculation for Equatorial Bulge}
According to the section \ref{descrip}, as the value of angular velocity increases the height of fluid layer ($m$) along $Z-$ axis will decrease resulting an equatorial bulge along $XY$ plane. Let the decrease of fluid layer along $Z-$axis be $h$ and Let the value of the bulge be $n$.
\begin{figure}[ht]
    \centering
    \includegraphics[scale=0.6]{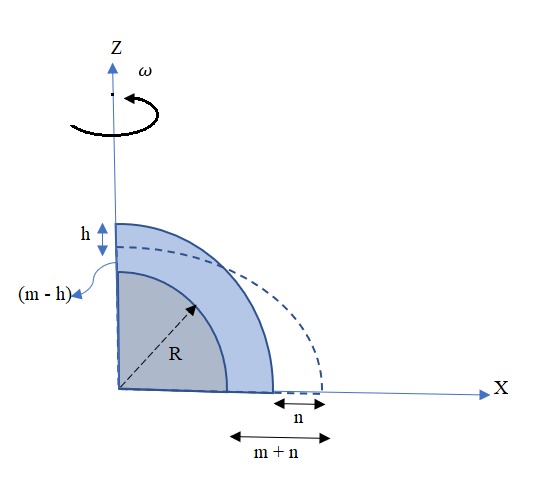}
  \caption{Quadrant Extension due to Rotation of the System}
    \label{fig:my_label}
\end{figure}\\
\\
Since the volume of the fluid is constant throughout the process, to calculate the value of the bulge we are going to conserve the volume of initial state and final state.\\
Let the volume of liquid at initial state be $V_1$ and that at final state be $V_2$, so we have:
\begin{align*}&~~~~~~ V_1 = \frac{4}{3}\pi(R+m)^3-\frac{4}{3}\pi R^3\\
&~~~~~~ V_2 = \frac{4}{3}\pi(R+m+n)^2(R+m-h)-\frac{4}{3}\pi R^3
\end{align*}
Since $V_1 = V_2$, we can write:
\begin{align*}&~~~~~~~~
\frac{4}{3}\pi(R+m)^3-\frac{4}{3}\pi R^3 = \frac{4}{3}\pi(R+m+n)^2(R+m-h)-\frac{4}{3}\pi R^3\\
&~~~~~~~~~~~~~~~~~\implies (R+m)^{3}=(R+m+n)^{2}(R+m-h)
\end{align*}
Since in the final state we consider the fluid layer along $Z-$ axis to just touches the metal ball surface, we have $m = h$. 
\begin{align*}&~~~~~~~~~~~~(R+m)^{3} = (R+m+n)^{2}R
\end{align*}
\begin{align}&\implies \boxed{n=(R+m)\left(\sqrt{1+\dfrac{m}{R}}-1\right)}
\label{eq1}
\end{align}
\subsection{Equation of the Spheroid}
\begin{figure}[ht]
    \centering
    \includegraphics[scale=0.5]{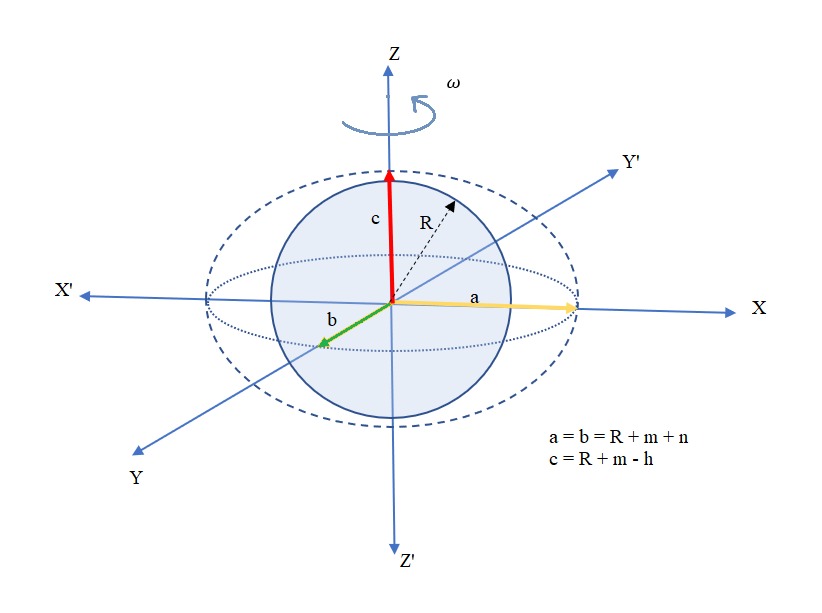}
  \caption{Fluid profile of the System rotating at any arbitrary $\omega$ }
    \label{fig:my_label1}
\end{figure}
So Finally we have the lengths of all semi major axes of the spheroid\\
$a = b = R+m+n= (R+m)\left(\sqrt{1+\dfrac{m}{R}}\right)$ \& $c = R$.\\
So Equation of the spheroid will be: 
\begin{equation}
\boxed{\dfrac{R(x^2+y^2)}{(R+m)^{3}}+\dfrac{z^2}{R^2}=1}   
\end{equation}
\section{Pressure Potential and containing force}
\label{ppcf}
\subsection{Derivation of Central Fluid Pressure}
Consider the equilibrium of an incompressible fluid with a constant angular velocity $\omega$ in some inertial frame of reference. According to standard Newtonian Dynamics the force balance equation takes form 
\begin{equation}
 0 = \nabla p + \rho \nabla \Psi +\rho \vec{\omega}\times(\vec{\omega}\times \vec{r})    
\end{equation}
Where $\nabla p$ is the static fluid pressure, $\rho$ is the mass density and in our case $\Psi$ is the potential for containing force.
So the Equation can be modified into 
\begin{equation}
    0 =  \nabla p + \rho \nabla (\Psi+\Psi')
\end{equation}
 Where $\Psi'$ is the centrifugal potential, $\Psi'= -\dfrac{\omega(x^2+y^2)}{2}$
 So, the equation becomes
 \begin{equation}
 \label{5}
     P = P_{0}-\rho \Psi + \dfrac{\rho \omega^{2}(x^2+y^2)}{2} 
 \end{equation}
 Where $P_{0}$ is Central Fluid Pressure and the pressure at the  outer boundary of the fluid surface must be zero, otherwise there would be a force imbalance across the boundary. Hence $P = 0$ which implies 
 \begin{equation}
    \boxed{P_{0} = \rho \Psi - \dfrac{\rho \omega^{2}(x^2+y^2)}{2}} 
 \end{equation}
 \subsection{Derivation of Containing Force}
 \label{cf}
 Now as we know, the effect of gravitational potential is much minimal than another potential taken under consideration in the equation for Central fluid pressure. The cause of this potential can be explained by presence of another form of force different from the gravitational one. This force in some form, is related to the surface formation of the liquid, which also includes surface tension ($\gamma$) of the fluid in the potential equation. Now surface tension of the fluid leads to \textbf{The Laplace pressure}, which is the pressure difference between the inside and the outside of a curved surface that forms the boundary between a gas region and a liquid region, and is given by:
 $$\Delta P = P_{\text{inside}}-P_{\text{outside}} = \gamma\left(\dfrac{1}{R_{1}}+\dfrac{1}{R_{2}}\right)$$ 
 Where $R_{1}$ and $R_{2}$ are the principle radii curvature of the fluid layer. And hence, pressure due to the potential for containing force can be represented by:
 \begin{align*}
     &\rho \Psi = \gamma\left(\dfrac{1}{R_{1}}+\dfrac{1}{R_{2}}\right)
 \end{align*}
 So the final Equation for containing force potential is: 
 \begin{equation}
 \label{7}
     \boxed{\Psi = \dfrac{\gamma}{\rho}\left(\dfrac{1}{R_{1}}+\dfrac{1}{R_{2}}\right)}
 \end{equation}
For General Ellipsoid, the principle radii of curvature depend on the Mean Curvature($H$) and the Gaussian Curvature($K$)~\cite{curvature} . Hence, $R_{1}= \dfrac{1}{H-\sqrt{H^{2}-K}}$ and $R_{2}=\dfrac{1}{H+\sqrt{H^{2}-K}}$.\\ 
That means equation \ref{7} can be modified as: 
\begin{align*}&~~~~~~ \Psi(x,y,z) = \dfrac{\gamma}{\rho}\left(\dfrac{1}{R_{1}}+\dfrac{1}{R_{2}}\right)\\
& \implies  \Psi(x,y,z) =\dfrac{\gamma}{\rho}\left(H-\sqrt{H^2-K}+H+\sqrt{H^2-K} \right)\\
& \implies \Psi(x,y,z) = \dfrac{2H\gamma}{\rho}\\
&\text{Now, As we know that}~~H=\dfrac{{\left|x^2+y^2+z^2-a^2-b^2-c^2\right|}}{2(abc)^{2}\left(\dfrac{x^2}{a^4}+\dfrac{y^2}{b^4}+\dfrac{z^2}{c^4}\right)^{\frac{3}{2}}}\\
& \implies \Psi(x,y,z) = \dfrac{\gamma}{\rho}\left\{\dfrac{{\left|x^2+y^2+z^2-a^2-b^2-c^2\right|}}{(abc)^{2}\left(\dfrac{x^2}{a^4}+\dfrac{y^2}{b^4}+\dfrac{z^2}{c^4}\right)^{\frac{3}{2}}}\right\} 
    \end{align*}
 So we get the general form of potential;
 \begin{equation}
     \boxed{\Psi(x,y,z) = \dfrac{\gamma}{\rho}\left\{\dfrac{{\left|x^2+y^2+z^2-a^2-b^2-c^2\right|}}{(abc)^{2}\left(\dfrac{x^2}{a^4}+\dfrac{y^2}{b^4}+\dfrac{z^2}{c^4}\right)^{\frac{3}{2}}}\right\}}
 \end{equation}
 So, the General form of Containing force will be 
 \begin{align*}
    & F(x,y,z)=-\nabla \Psi = -\left(\dfrac{\partial \Psi}{\partial x}+\dfrac{\partial \Psi}{\partial y}+\dfrac{\partial \Psi}{\partial z}\right)
 \end{align*}
 Hence; 
\begin{equation}
\label{eq9}
\boxed{ F(x,y,z)= \mathlarger{\mathlarger{\sum}}_{cyc~(a,b,c,x,y,z)}   \dfrac{\gamma}{\rho}\left\{\dfrac{3x\left|x^2+y^2+z^2-a^2-b^2-c^2\right|}{a^6b^2c^2\left(\frac{x^2}{a^4}+\frac{y^2}{b^4}+\frac{z^2}{c^4}\right)^\frac{5}{2}}-\dfrac{2x\left(x^2+y^2+z^2-a^2-b^2-c^2\right)}{a^2b^2c^2\left(\frac{x^2}{a^4}+\frac{y^2}{b^4}+\frac{z^2}{c^4}\right)^\frac{3}{2}\left|x^2+y^2+z^2-a^2-b^2-c^2\right|}\right\}}
\end{equation}
Now to analyse the potential and the containing force we only consider it's contribution along  X axis. So the simplified equations for potential and force will be:\\
$ \Psi(x,0,0) = \dfrac{\gamma a^4 \left|x^2-a^2\right|}{\rho x^3} \text{ and } F(x,0,0) = \dfrac{\gamma}{\rho}\left\{\dfrac{3a^4\left|x^2-a^2\right|}{x^4}-\dfrac{2a^4(x^2-a^2)}{x^2\left|x^2-a^2\right|}\right\} $.

\section{Angular Velocity for the required system}
\label{av}
We are calculating for which angular velocity ($\omega$) the fluid layer along $Z-$ axis just touches the surface of the metal sphere. To find the required angular velocity we have to use Conservation of Energy. Now since the rotation of the complete system is resulting in the change in the shape of the surface, hence total rotational energy can be considered to be transferred to the overall change in the surface energy due to changed shape of the fluid layer. Let $\gamma_{wa}$ and $\gamma_{ws}$ be the interfacial energy between water layer and the gas or air present outside and between water layer and solid metal surface respectively.Let $\Delta E_1$ and $\Delta E_2$ be the initial and final surface energies of the system respectively. Also, let the initial and final surface areas of the fluid layer be $S$ and $S'$ respectively. Then we can write: 
\begin{align*}
&~~~~~~\Delta E_1 = S\gamma_{wa} + 4\pi R^2\gamma_{ws}\\
&~~~~~~\Delta E_2 = S'\gamma_{wa} + 4\pi R^2\gamma_{ws}
\end{align*}
For conserving total energy of the system, we can say that total energy available initially will be equal to the total energy available finally with the system. Hence we can write:
\begin{align*}
&\text{Change in Rotational Kinetic Energy = Change in Surface Energy}\\
&~~~~~~~~~~~~~~~~~\implies~~~~~~~~~\frac{1}{2}I_{net}\omega^2 = \Delta E_2 - \Delta E_1\\
&~~~~~~~~~~~~~~~~~\implies~~~~~~~~~\frac{1}{2}I_{net}\omega^2 =(S'\gamma_{wa} + 4\pi R^2\gamma_{ws}) - (S\gamma_{wa} + 4\pi R^2\gamma_{ws})\\
&~~~~~~~~~~~~~~~~~\implies~~~~~~~~~ \frac{1}{2}I_{net}\omega^2 = (S'-S)\gamma_{ws}
\end{align*}
\begin{figure}[ht]
    \centering
    \includegraphics[scale=0.55]{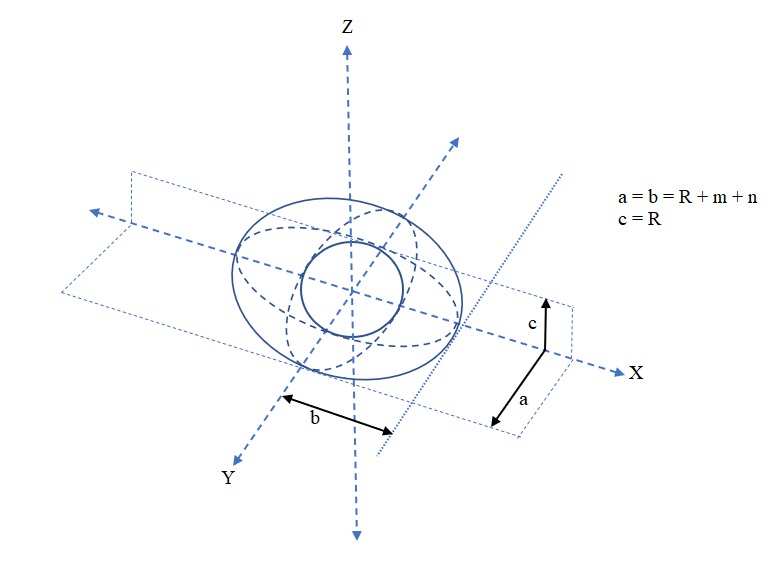}
  \caption{Spheroid Profile for Specific $\omega $ }
    \label{fig:my_label2}
\end{figure}
\pagebreak

Now, Moment of Inertia ($I_{net}$) of the overall system can be divided into three parts, i.e., moment of inertia of the overall spheroid($I_1$), moment of inertia of the cavity ($I_2$)(which is to be subtracted from that of the spheroid) and moment of inertia of the metal ball ($I_3$). Hence we can write:
\begin{align*}
&~~~~~~I_{net} = I_1 - I_2 + I_3\\
&\text{Now,}\\
&~~~~~~I_1 = \frac{2}{5}\rho \Bigg\{\frac{4}{3}\pi(R+m+n)^2R\Bigg\}(R+m+n)^2\\
&~~~~~~I_2 = \frac{2}{5}\rho\Bigg\{\frac{4}{3}\pi R^3\Bigg\}R^2\\
&~~~~~~I_3 = \frac{2}{5}\rho_m\Bigg\{\frac{4}{3}\pi R^3\Bigg\}R^2\\
&\text{Therefore we have,}\\
&~~~~~~~~~~I_{net} = \frac{8}{15}\rho \pi (R+m+n)^4 R - \frac{8}{15}\rho \pi R^5 + \frac{8}{15}\rho_m \pi R^5\\ 
&\text{Using Equation \ref{eq1}, we have} \\
&\implies I_{net} = \dfrac{8\pi}{15}R^5\left[\rho\left(1+\dfrac{m}{R}\right)^{6}+\rho_{m}-\rho\right] \end{align*}
Now the Surface area of the spheroid with equation $\dfrac{(x^2+y^2)}{a^2}+\dfrac{z^2}{c^2}=1$ is $S'= 2\pi\left[a^2+\dfrac{c^2}{\sin{\alpha}}\ln\left({\dfrac{1+\sin{\alpha}}{\cos{\alpha}}}\right)\right]$, where $\alpha = \arccos{\dfrac{c}{a}}$.\\
Now for our case we have $a =(R+m)\left(\sqrt{1+\dfrac{m}{R}}\right)$ and $c = R$, Hence ;
\begin{equation}
    S' = 2\pi\left[\dfrac{(R+m)^{3}}{R}+R^{2}\sqrt{\dfrac{(R+m)^{3}}{(R+m)^3-R^3}}\ln\left(\sqrt{\left(1+\dfrac{m}{R}\right)^{3}}+\sqrt{\left(1+\frac{m}{R}\right)^{3}-1}\right)\right]
\end{equation}
So using the value of $S'$ and $I_{net}$ we can find the value of $\omega$ as:
\begin{equation}
    \boxed{\omega = \sqrt{\dfrac{15\gamma_{wa}}{2R^3}\left\{\dfrac{\left(1+\dfrac{m}{R}\right)^{3}+\sqrt{\dfrac{(R+m)^{3}}{(R+m)^3-R^3}}\ln\left(\sqrt{\left(1+\dfrac{m}{R}\right)^{3}}+\sqrt{\left(1+\dfrac{m}{R}\right)^{3}-1}\right)-2\left(1+\dfrac{m}{R}\right)^{2}}{\rho\left(1+\dfrac{m}{R}\right)^{6}+(\rho_{m}-\rho)}\right\}}}
\end{equation}
Now for $m<<R$; we can approximate the value of $\omega$ by using Taylor's Series~\cite{taylor} up to $O(1)$ and from that we get the expression of $\omega$ as: 
\begin{equation}
    \omega =\sqrt{\dfrac{15\gamma_{wa}}{2R^3}\left\{\dfrac{\left(\sqrt{\dfrac{R}{3m}+1}\right)\ln\left(\sqrt{\left(1+\dfrac{3m}{R}\right)}+\sqrt{\dfrac{3m}{R}}\right)-\left(1+\dfrac{m}{R}\right)}{\left(\dfrac{6\rho m}{R}+\rho_{m}\right)} \right\}}
\end{equation}
\pagebreak
\section{Limiting Conditions}
\label{lc}
\begin{figure}[ht]
    \centering
    \includegraphics[scale=0.5]{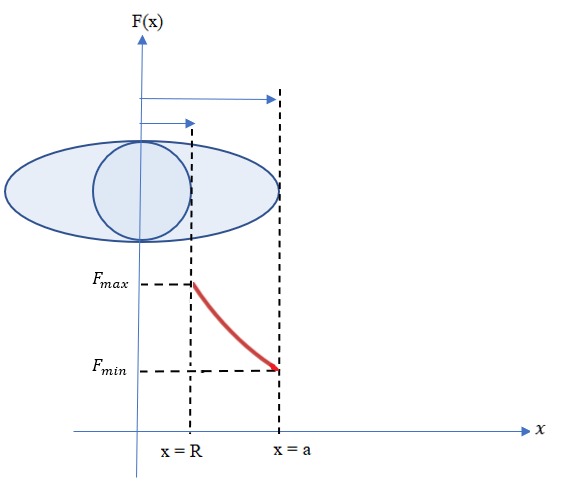}
  \caption{Variation of Containing Force $F(x)$ along X-axis}
    \label{fig:force graph}
\end{figure}

The simplified expression for Force in equation \ref{eq9}  can be plotted with respect to distance measured along $X$-axis as shown in Figure \ref{fig:force graph}. Now as we can see that we had restricted the plot of $F(x)$ from the radius of the metal sphere $R$, up to the semi- major axis along $X$ axis or $Y$ axis, $a = (R+m)\sqrt{1+\dfrac{m}{R}}$, which corresponds to the force values $F_{max}$ and $F_{min}$ respectively. The restriction comes with the reason that the considered containing force is limited to a particular value of $x$ beyond which, the magnitude of force becomes inconsistent with the physical system as fluid is present from the surface of the metal sphere up to the liquid and air interface.
\begin{equation*}
\text{Now,}~~F(x) = \frac{\gamma}{\rho}\left[\dfrac{3a^4(a^2-x^2)}{x^3}+\dfrac{2a^4}{x^2}\right]~~~  (\text{When,}~ 0\leq x \leq a)   
\end{equation*}
The force at $x = a$ is given by $F_{min}= F(a) = \dfrac{ 2a^2\gamma}{\rho}$ . This agrees with the fact that for the system being static or without any rotation, this force is responsible for bringing the surface of the liquid to the equilibrium position, i.e. into the spherical shape. Now during rotation, this force is balanced by the centrifugal force acting per unit mass along the $X$ direction. Using this concept we can find a bound on the angular velocity ($\omega$) for which the water layer remains in shape of the spheroid without spilling out from the system.
\begin{align*} &F_{min} = F(a) = \dfrac{ 2a^2\gamma}{\rho} \geq \omega^{2}a\\
&\implies~~~~~~\boxed{ \omega \leq \sqrt{\dfrac{2 a \gamma}{\rho}}} 
\end{align*}
Now if we give  $\omega > \sqrt{\dfrac{2 a \gamma}{\rho}}$, the water will start spilling out from the outer surface as the force at the boundary is minimum.
\pagebreak
\begin{figure}[ht]
    \centering
    \includegraphics[scale=0.5]{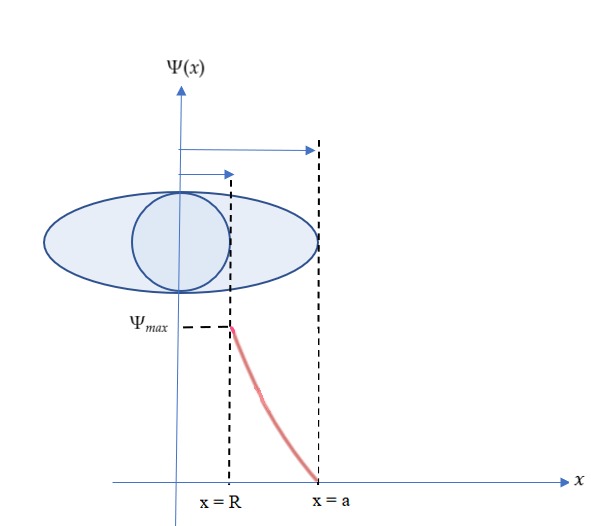}
  \caption{Variation of Containing Potential $\Psi(x)$ along X-axis}
    \label{fig:psi graph}
\end{figure}

Similarly, with same conditions, we can define the containing potential of the system along $X$ direction from radius $R$ to the value $a$ as shown in Figure \ref{fig:psi graph} . The potential trend also have physical significance, since potential at the surface of the spheroid surface of the liquid is zero. Now with regards to the $F_{max}$, this force is balanced by the surface of the metal sphere, since the metal sphere is rigid and is not deformable.

\section{Conclusion}
\label{con}
We are able to highlight certain points which may not seem possible to replicate on earthly conditions, by removing effects on gravity but keeping other conditions around it as such. Under theoretical considerations, we were able to predict specific aspects like how much angular velocity is required in order to get the desired shape of the liquid drop with condition of having very little trace of liquid at region along $Z$ axis, also called polar region. We have to introduce the concept of containing force, which is other than gravitational force present between the particles of the liquid and the metal sphere. The reason behind introduction of such concept within our system is also explained as we are not considering very massive system like that of a planet and its atmospheric layer. We also explained how the containing force is helpful for explaining the presence of a potential, which is different from gravitational potential as explained in the \textbf{Maclaurin Spheroid} \cite{maclaurin}. If we consider the gravitational force and corresponding gravitational potential, the expression for angular velocity ($\Omega$) is already derived as:
$${\displaystyle {\frac {\Omega ^{2}}{\pi G\rho }}={\frac {2{\sqrt {1-e^{2}}}}{e^{3}}}(3-2e^{2})\sin ^{-1}e-{\frac {6}{e^{2}}}(1-e^{2}),\quad e =\text{Eccentricity of the spheroid}} $$
Further we proceed to calculate the restriction on the angular velocity beyond which, the centrifugal force will tend to spill the fluid from the system, leaving all the expressions explaining the system to be inconsistent.

The system we tried to explain in the paper seems to have least amount of applications in current physics and engineering fields, but it has the potential to explain about the considerations related to outer space machineries with special focus on lubrication systems. 
\section{Acknowledgement}
\label{ack}
RT, SM and KP would like to thank School of Physical Sciences (SPS) and Academic Section of NISER, Bhubaneswar where they got the opportunity to interact with wonderful members and professors who helped them a lot with the  basics of  Fluid Dynamics. They also acknowledge the support of their parents who constantly kept them motivated throughout the project during the COVID-19 Pandemic and didn't let their morale down.

\end{document}